\documentclass[a4paper,11pt]{article}
\usepackage{jheppub}
\usepackage{lineno}
\usepackage{amsfonts,amssymb,epsfig,amsmath,mathtools}
\usepackage{color}
\usepackage{comment}
\usepackage{url}
\usepackage{mathrsfs}
\usepackage{graphicx}
\usepackage{subcaption}
\usepackage{hyperref}

\newcommand{\be}{\begin{eqnarray}}
\newcommand{\ee}{\end{eqnarray}}

\newcommand{\bn}{\begin{enumerate}}
\newcommand{\en}{\end{enumerate}}

\title{\boldmath 
\vspace*{-0.3cm}
Probing Earth-Bound Dark Matter \\  with Nuclear Reactors 
\vspace{-0.2cm}
}

\author[a,b]{Yohei Ema}
\author[a,b]{Maxim Pospelov}
\author[b,c]{Anupam Ray}
\affiliation[a]{William I. Fine Theoretical Physics Institute, School of Physics and Astronomy, \\ University of Minnesota, Minneapolis, MN 55455, USA}
\affiliation[b]{School of Physics and Astronomy, University of Minnesota, Minneapolis, MN 55455, USA}
\affiliation[c]{Department of Physics, University of California Berkeley, Berkeley, California 94720, USA}

\emailAdd{ema00001@umn.edu}
\emailAdd{pospelov@umn.edu}
\emailAdd{anupam.ray@berkeley.edu}

\abstract{Strongly-interacting dark matter can be accumulated in large quantities inside the Earth, and for dark matter particles in a few GeV mass range, it can exist in large quantities near the Earth's surface. We investigate the constraints imposed on such dark matter properties by its upscattering by fast neutrons in nuclear reactors with subsequent scattering in nearby well-shielded dark matter detectors, schemes which are already used for searches of the coherent reactor neutrino scattering. We find that the existing experiments cover new parameter space on the spin-dependent interaction between dark matter and the nucleon. Similar experiments performed with research reactors, and lesser amount of shielding,  may provide additional sensitivity to strongly-interacting dark matter.}
\preprint{\parbox{3.5 cm}{UMN-TH-4312/24\\FTPI-MINN-24-04\\N3AS-24-002}}

\begin{document}
\maketitle
\flushbottom
\section{Introduction}
\label{sec:introduction}

One of the first signs that the Standard Model of particles and fields (SM) needs to be extended comes from cosmology and astrophysics. Large classes of fairly diverse data point to the existence of the energy-density component in the Universe that can cluster and form cosmic structures. This component, a manifestation of some beyond-SM physics (BSM), is commonly referred to as dark matter (DM). Despite overwhelming evidence for the DM existence, its identity remains a mystery. A number of important classes of candidates were explored both experimentally and theoretically over the past decades. These include weakly interacting massive particles, or WIMPs, as well as super-weakly interacting bosonic fields such as axions. Despite tremendous efforts spent both theoretically and experimentally, no convincing signals of these particles have been found. In this paper, we will concentrate on one type of dark matter that interacts stronger than a WIMP, but nevertheless stays elusive to most of the terrestrial experiments.

If cross sections of DM scattering on atoms are very large, DM particles can quickly thermalize in Earth's atmosphere, rock and soil, making it energetically sub-threshold for most of the direct detection detectors, 
\begin{equation}
	E_{\chi, \rm kin} \to kT_{\rm Earth} < E_{\rm threshold}.
\end{equation}
Indeed, the kinetic energy of $kT_{\rm Earth}\sim 0.03$\,eV is typically far below the sensitivity of modern devices.  It is often the case that if strongly-interacting particles {\em saturate} the entire DM abundance, the limits from the balloon and surface direct detection experiments are too strong (see {\em e.g.} Ref. \cite{Collar:2018ydf,Bramante:2022pmn}). 
However, for a fractional DM abundance, characterized by $f_\chi$, $f_\chi \leq 1$, large interaction cross sections are not excluded. (See, however, an opposing opinion \cite{Xu:2020qjk} claiming that for some limited range of masses and cross sections $f_\chi=1$ is still allowed.)

There have been numerous ideas on how to probe $f_\chi \ll 1, l_{\rm coll} \ll R_\oplus$ models of DM, where $l_{\rm coll}$ stands for a typical DM-atom scattering length. If thermalization happens beyond 100\,m\,--\,1\,km length scale, then usual direct detection (DD) constraints apply up to $\mathcal{O}(1)$ factor. If however the thermalization is rapid, then one has to invent alternative probes. Several ideas have been exploited so far:
\begin{enumerate}
	\item Effects of thermalized DM on collection of SM particles ({\em e.g.} particles in a beam, or very cold atoms such as in liquid helium) \cite{Neufeld:2018slx,Xu:2021lmg,Farrar:2022mih}. 
	
	\item Existence of very fast component of DM that has more penetrating power (such as fast DM component arising from collisions with cosmic rays \cite{Bringmann:2018cvk,Ema:2018bih,Cappiello:2019qsw}). 
	
	\item Annihilation to visible and neutrino modes inside the Earth~\cite{McKeen:2023ztq,Pospelov:2023mlz}. 
	
	\item De-excitation of nuclear isomers that is enhanced because of the relatively large momentum exchange, even with thermalized DM \cite{Pospelov:2019vuf,Lehnert:2019tuw,Majorana:2023ecz}, and power measurement using novel quantum devices \cite{Das:2022srn}. 
	
	\item Possibility to use laboratory sources to {\em accelerate} DM and study its subsequent scattering. In particular, the use of underground accelerators was entertained in Refs. \cite{Pospelov:2020ktu,McKeen:2022poo}. 
\end{enumerate}

In this paper, we explore the idea very similar to the last one on this list. Namely, we analyze the {\em existing} constraints from coupling of nuclear reactors with well-shielded nearby DD devices. Experimentally, this is not a hypothetical scheme, but a mature effort 
directed to detect the coherent neutrino scattering on nuclei 
(CE$\nu$NS)  at $E_\nu < $\,few\,MeV (see e.g. Refs.~\cite{CHANDLER:2022gvg,Abdullah:2022zue}). The essence of our idea is as follows: nuclear reactors, while operating, constantly generate the flux of MeV-energy neutrons. Prior to/during the collisions with a moderator ({\em e.g.} $p$ in ${\rm H_2O}$), neutrons have a chance of interacting with DM, thereby accelerating it to velocities that are much larger than thermal. Subsequent scattering with an atom inside the detector can be visible. Thus, we suggest to use the nuclear reactors as the most powerful ``$\mathcal{O}({\rm MeV})$ neutron beam dumps".

We find that the existing experiments at commercial nuclear reactors provide additional constraints on thermalized fraction of DM that are almost competitive with the existing DD constraints in the spin-independent (SI) interaction case, and probe a new parameter region
not explored by the existing DD experiments in the spin-dependent (SD) interaction case, respectively.
In the future, the sensitivity to thermalized DM can be further improved, providing an additional physics motivation for ``coupling" DD devices to nuclear reactors. We note that the research nuclear reactors, that are less powerful than the commercial ones, require less shielding around a neutrino/DM detector, and therefore could provide additional constraints in the large cross section/low scattering range regime. 

This paper is organized as follows. In the next section we review the accumulation and distribution of strongly-interacting DM in the Earth. We calculate the velocitized DM flux generated by the up-scattering on neutrons. In Section~\ref{sec:constraints} 
we analyze possible signal, solving the problem of DM slow-down due to multiple collisions via numerical simulations. We reach our conclusions in Section~\ref{sec:summary}.

\section{Earth-bound dark matter flux accelerated by nuclear reactor}
\label{sec:DMflux}

The Earth is exposed to a continuous flux of DM particles,
and a component of the DM strongly interacting with the SM particles, if exits,
can be efficiently captured by the Earth.
As a result, the number density of the Earth-bound DM particles greatly exceeds the standard halo DM number density,
and yet these are harder to detect by the standard DM direct detection experiments due to their small kinetic energy
after thermalization.
In this regard, nuclear reactors can play an interesting role
as neutrons inside the reactors can upscatter the Earth-bound DM particles, 
enhancing their kinetic energy to be observed at nearby detectors.

To study the nuclear-reactor-generated flux of non-thermal DM, we need to combine the following ingredients: {\em i.} accumulation of dark matter inside the Earth, {\em ii.} upscattering of DM on fast neutrons inside the reactor, {\em iii.} subsequent propagation through shielding and eventual scattering inside a DM detector. In this section, we address these topics in turn.

\subsection{Earth-bound dark matter}
\label{subsec:EarthDM}
A strongly-interacting dark matter component (which we call $\chi$), owing to their interactions with the Earth-nuclei, efficiently accumulates inside the Earth-volume. Their time evolution inside the Earth-volume (assuming no annihilation among them) 
is determined by the balance between their capture ($\Gamma_\mathrm{cap}$) and thermal evaporation ($\Gamma_\mathrm{evap}$)
\begin{align}
	\frac{d N_\chi}{dt} = \Gamma_\mathrm{cap}
	- \Gamma_\mathrm{evap} N_\chi\,.
	\label{eq:dNchidt}
\end{align}
In the following, we briefly discuss both of these rates.

Starting with $\Gamma_\mathrm{cap}$, we first define the maximal capture rate as geometric capture rate $(\Gamma_\mathrm{geo})$ which occurs when all of the transiting DM particles through the Earth get captured. Of course, 
the capture rate, $\Gamma_\mathrm{cap}$, is a certain fraction of the geometric capture rate which crucially depends on the DM mass as well as DM-nucleon scattering cross section. In the optically thick regime, i.e., for large DM-nucleon scattering cross section, this fraction, commonly known as capture fraction $(f_\mathrm{cap})$, can be quite significant. For heavier DM $(m_{\chi} \gg m_A$ with $m_A$ is the typical nuclear mass), it can even reach unity, whereas, for GeV-scale DM, $f_\mathrm{cap} \sim 0.1$~\cite{Bramante:2022pmn}. 
We use recent numerical simulations~\cite{Bramante:2022pmn} to estimate the value of $f_\mathrm{cap}$ which agrees reasonably well with the previous analytical estimate~\cite{Neufeld:2018slx}. To summarize, we use the following capture rate for the optically thick
regime
\begin{align}
	\Gamma_\mathrm{cap} = f_\mathrm{cap}\times \Gamma_\mathrm{geom}
	= f_\mathrm{cap} \times \sqrt{\frac{8}{3\pi}}\frac{f_\chi \rho_\mathrm{DM} v_\mathrm{gal}}{m_\chi}\times 
	\pi R_\oplus^2,
\end{align}
where $f_\chi = \rho_\chi/\rho_\mathrm{DM}$ is the energy fraction of $\chi$ with $\rho_\mathrm{DM}$ being the local DM energy density,
$m_\chi$ is the mass of $\chi$,  $v_\mathrm{gal}$ is the average DM velocity inside the galaxy,
and $R_\oplus$ is the Earth radius.
We take $\rho_\mathrm{DM} = 0.4\,\mathrm{GeV}\mathrm{cm}^{-3}$, $v_\mathrm{gal} = 220\,\mathrm{km/s}$,
and $R_\oplus = 6371\,\mathrm{km}$ in our numerical computation.

Thermal evaporation, which is particularly important for light DM, occurs when the DM particles obtain too much kinetic energy via thermal kicks from the stellar nuclei and overcome the escape velocity of the stellar object.  If the collision length of the DM particles is significantly smaller than the size of the Earth, which is of primary interest here, the evaporation effectively occurs from the “last scattering surface”, and we estimate the evaporation rate by adopting the Jeans’ formalism~\cite{Neufeld:2018slx}
\begin{align}
	\Gamma_\mathrm{evap} = G_\chi\left(R_\mathrm{LSS}\right) \times \frac{3R_\mathrm{LSS}^2}{R_\oplus^3}
	\times \frac{v_\mathrm{LSS}^2 + v_\mathrm{esc}^2}{2\pi^{1/2}v_\mathrm{LSS}}
	\exp\left(-\frac{v_\mathrm{esc}^2}{v_\mathrm{LSS}^2}\right),
\end{align}
where $R_\mathrm{LSS}$ $(v_\mathrm{LSS})$ denotes the radius (thermal velocity) at the last scattering surface. The thermal velocity at the last scattering surface $(v_\mathrm{LSS})$ is related to the temperature at the last scattering surface as $v_\mathrm{LSS} = \sqrt{2T_\mathrm{LSS}/m_\chi}$. We obtain $R_\mathrm{LSS}$ via
\begin{equation}
	\int_{R_{\rm LSS}}^{\infty} dr \sum_j \sigma_{\chi j} n_j(r) =1\,,
	\label{eq:RLSS}
\end{equation}
where $\sigma_{\chi j}$ denotes the scattering cross-section between DM and the $j$-th nuclei and $n_j(r)$ denotes the number density of the $j$-th nuclei. For the density profile of the Earth, we use Preliminary Reference Earth Model~\cite{Dziewonski:1981xy}, and for the chemical compositional profile, we use Table I of Ref.~\cite{Bramante:2019fhi}.\footnote{
	Most elements inside the Earth do not have a nuclear spin, and relatively rare isotopes such as $^{29}$Si and $^{27}$Al provide the dominant contribution to the last scattering surface inside the crust in the SD case.
	To evaluate Eq.~\eqref{eq:RLSS}, we have adopted the rescaling~\eqref{eq:sigmaSD_rescale} given below (with $m_\mathrm{Si}$ replaced by each element and the spin contents given in~\cite{Bramante:2022pmn}).
} For the density, temperature, and compositional profile of the Earth's atmosphere, we use NRLMSISE-00 model~\cite{2002JGRA..107.1468P}.
Note that, for our parameter range of interest, $R_{\rm LSS}$ resides either at the Earth's crust or slightly above the crust, i.e., $R_\mathrm{LSS} \simeq R_\oplus$~\cite{Neufeld:2018slx}. To be more specific, the height of the atmosphere, relevant in our analysis, is at most $\sim 80\,\mathrm{km}$, and is therefore negligible compared to the Earth's radius $R_\oplus = 6371\,\mathrm{km}$, justifying $R_\mathrm{LSS} \simeq R_\oplus$.

Finally, $G_\chi (r) = V_\oplus n_\chi(r)/ N_{\chi}$ represents the spatial distribution of the DM particles within the Earth-volume.
This is governed by the Boltzmann equation that combines the effects of gravity, concentration diffusion, and thermal diffusion~\cite{Gould:1989hm,Leane:2022hkk}
\begin{align}
	\label{diff}
	\frac{\nabla n_{\chi}(r) }{n_{\chi}(r)}+ \left(\kappa +1 \right) 	\frac{\nabla T(r) }{T(r)}+\frac{m_{\chi} g(r)}{k_BT(r)} = 0\,.
\end{align}
In the above equation, we have used the
hydrostatic equilibrium criterion as the diffusion timescales for the $\chi$ particles are short as compared to the other relevant timescales.  $\kappa \sim -1/\left[2(1+m_{\chi}/m_{A})^{3/2}\right]$ denotes the thermal
diffusion co-efficient~\cite{Leane:2022hkk}, $g(r)$ is the acceleration due to gravity, and $T(r)$ denotes the temperature profile of the Earth~\cite{https://doi.org/10.1002/2017JB014723}. For a uniform DM distribution inside the Earth, $G_\chi(r) = 1$. We note that, in our parameter range of interest, the thermal diffusion co-efficient $\kappa$, which depends on the mass of the nuclei $(m_A)$, remains almost constant for a given DM mass. Given such a weak dependence of $\kappa$ with $m_A$, we have taken $m_A = 16m_n$ ($m_A = 27m_n$ or $m_A = 29m_n$) for spin-independent (spin-dependent) interactions, while solving the Boltzmann equation to obtain $G_\chi$. Whereas, when we obtain the last scattering surface in Eq.~\eqref{eq:RLSS}, which is crucial to estimate the evaporation rate, we have included the more realistic chemical composition of the Earth.

By solving Eq.~\eqref{eq:dNchidt} with the initial condition $N_\chi = 0$ at $t = 0$, we obtain
\begin{align}\label{evap2}
	N_\chi (t_{\oplus}) = \frac{\Gamma_\mathrm{cap}}{\Gamma_\mathrm{evap}}
	\left(1-e^{-t_\oplus \Gamma_\mathrm{evap}}\right),
\end{align}
where $t_\oplus \simeq 1.4\times 10^{17}\,\mathrm{sec}$ 
is the age of the Earth. When evaporation is negligible, Eq.~\eqref{evap2} simply reduces to $N_\chi (t_{\oplus}) \approx t_{\oplus} \Gamma_\mathrm{cap}$. DM number density at the Earth's surface, where nuclear reactors are, is therefore given by
\begin{align}
	n_\chi (R_{\oplus}) = G_\chi (R_{\oplus}) \times \frac{N_\chi}{V_\oplus}
	= G_\chi (R_{\oplus}) \frac{\Gamma_\mathrm{cap}}{\Gamma_\mathrm{evap} V_{\oplus}}
	\left(1-e^{-t_\oplus \Gamma_\mathrm{evap}}\right)
\end{align}
In the light mass region, $m_\chi \lesssim 1\,\mathrm{GeV}$,
this is suppressed by the efficient evaporation, while in the heavy mass region, $m_\chi \gtrsim 3\,\mathrm{GeV}$,
the normalized surface density $G_\chi$ starts to drop since most of the DM sinks to the center of the Earth.\footnote{
	For the heavy mass region, the DM number density is enhanced due to 
	the traffic jam effect~\cite{Pospelov:2019vuf,Lehnert:2019tuw,Pospelov:2020ktu}.
	We do not include it in our analysis since it does not significantly enhance the sensitivity for our region of interest.

}
As a result, the DM number density at the location of the reactors takes its maximal value at $m_\chi \sim 2\,\mathrm{GeV}$.
\subsection{Dark matter acceleration at nuclear reactor}
Nuclear reactors utilize a fission chain reaction to extract the nuclear energy.
Each fission reaction produces a few energetic neutrons, $E_n \gtrsim \mathrm{MeV}$,
and the reactors slow them down, or moderate them, as the fission cross section 
is greatly enhanced at the lower neutron energy.
To explore BSM physics, one can thus think of the nuclear reactor 
as a neutron beam dump experiment;
energetic neutrons are injected into a moderator (typically, water or heavy water)
and potentially produce and/or upscatter BSM particles before losing the kinetic energy.
In particular, in the case of our interest, the reactor neutrons
can upscatter DM particles accumulated inside the Earth,
transferring them sufficient kinetic energy to be observable at detectors near the reactors.

$^{235}$U emits about 2.5 neutrons and $200\,\mathrm{MeV}$ per fission, released mostly as kinetic energy of the fission products.
From this information, we can estimate the total number of neutrons inside the nuclear reactor per time as
\begin{align}
	\frac{d N_n}{dt} \simeq 7.8 \times 10^{19}\,\mathrm{sec}^{-1}\times \left(\frac{P}{1\,\mathrm{GW}}\right),
\end{align}
where $P$ is the thermal power of the nuclear reactor, and we normalize it on a typical power of a large commercial reactor.
One can obtain similar estimates for $^{238}$U and $^{239}$Pu.
We have checked that the same estimate reproduces the antineutrino flux 
at the location of the CONUS detector~\cite{CONUS:2020skt,CONUS:2021dwh}.
The fission neutron spectrum is well-fitted by the so-called Watt distribution,
and the normalized spectrum is given (for $^{235}$U) as~\cite{Cranberg:1956zz}
\begin{align}
	\frac{1}{N_n}\frac{dN_n}{dE_n} = \frac{0.453}{\mathrm{MeV}}\,e^{-E_n/0.965\,\mathrm{MeV}}
	\sinh\left[\sqrt{2.29 E_n/\mathrm{MeV}}\right].
\end{align}

We treat it as the neutron injection spectrum, and calculate the resulting flux of the DM, while taking into account the elastic scattering of neutrons by hydrogen in water. (We will not consider heavy water reactors, noting that one can easily extend the same treatment to ${\rm D_2O}$).
The probability of a single neutron upscattering the DM while traveling the distance $dx$ is
\begin{align}
	n_\chi \sigma_{\chi n} dx = n_\chi \int d{E_\chi}\frac{d\sigma_{\chi n}}{dE_\chi} dx,
\end{align}
where $n_\chi$ is the DM number density at the reactor which we compute by following Sec.~\ref{subsec:EarthDM},
$\sigma_{\chi n}$ is the cross section between the neutron and DM,
and $E_\chi$ is the energy of the final state DM.
In our study, we consider both SI and SD interactions, 
$\sigma_{\chi n}^{\mathrm{SI}}$ and $\sigma_{\chi n}^{\mathrm{SD}}$,
and the discussion here equally applies to both cases.
The spectrum of the DM from a single neutron, before sufficiently moderated, is given by
\begin{align}
	\frac{d\bar{N}_\chi}{dE_\chi}(E_n, E_\chi) &= n_\chi \int_{0}^{x_\mathrm{th}(E_n)}dx
	\frac{d\sigma_{\chi n}}{dE_\chi}(E(x),E_\chi)
	= n_\chi \int_{E_\mathrm{th}}^{E_n} \frac{dE}{-dE/dx}\frac{d\sigma_{\chi n}}{dE_\chi}(E,E_\chi),
\end{align}
where $E(x)$ is the neutron energy after traveling the distance of $x$ with $E(0) = E_n$ 
being the initial neutron kinetic energy, 
$-dE/dx$ is the energy loss rate of the neutron inside the reactor,
and $E_\mathrm{th}$ ($x_\mathrm{th}$) is the threshold energy (travel distance) of the neutron we follow,
which depends on the DM signal of our interest.
Number density of dark matter $n_\chi$ is to be taken as $n_\chi (R_{\oplus})$ calculated in Sec.~\ref{subsec:EarthDM}.
The energy loss function of neutrons, $-dE/dx$, is well studied in the context of, \emph{e.g.}, nuclear reactor engineering,
but we may take a simplified treatment and estimate it as
\begin{align}
	-\frac{dE}{dx} = E_\mathrm{loss} n_p \sigma_{pn},
\end{align}
where $n_p$ is the number density of the target particle (or the proton), $\sigma_{pn}$ is the proton-neutron cross section, 
and $E_\mathrm{loss} = E/2$ is the mean energy loss per scattering,
assuming that the moderator is composed of water.\footnote{
	Oxygen nucleus is heavier than the neutron, and has smaller elastic scattering cross section on $n$, and hence is less important as a moderator.
}
The neutron-proton cross section is well-known,
and we include it numerically in our computation.
Combining these formulas, the DM number spectrum at the reactor is estimated as
\begin{align}
	\frac{dN_\chi}{dE_\chi} = \int dE_n \frac{dN_n}{dE_n}\frac{d\bar{N}_\chi}{dE_\chi}(E_n, E_\chi)
	= n_\chi 
	\int dE_n \frac{dN_n}{dE_n}\int_{E_\mathrm{th}}^{E_n} \frac{dE}{-dE/dx}\frac{d\sigma_{\chi n}}{dE_\chi}(E,E_\chi).
\end{align}
Notice that this formula closely resembles those appearing in the context of beam dump experiments.
In the following, we assume that the DM-neutron scattering is isotropic, which results in
\begin{align}
	\frac{d\sigma_{\chi n}}{dE_\chi}(E, E_\chi) = \frac{\sigma_{\chi n}}{\beta_{\chi n} E}
	\Theta\left(\beta_{\chi n} E - E_\chi\right),
	\quad
	\beta_{\chi n} = \frac{4 m_\chi m_n}{(m_\chi + m_n)^2},
\end{align}
where $\Theta$ is the Heaviside step function
and we take the non-relativistic limit of the neutron and the DM.
The accelerated DM spectrum at the source is now simplified as
\begin{align}
	\frac{dN_\chi}{dE_\chi} = \frac{2}{\beta_{\chi n}}\int_{E_\chi/\beta_{\chi n}}^{\infty} 
	dE_n \frac{dN_n}{dE_n}
	\int^{E_n}_{E_\chi/\beta_{\chi n}} \frac{dE}{E^2}\frac{n_\chi \sigma_{\chi n}}{n_p \sigma_{pn}(E)}.
\end{align}
Plugging in benchmark values, we obtain
\begin{align}
	\frac{d^2 N_\chi}{dt dE_\chi} 
	\simeq \frac{1.6\times 10^{6}\,\mathrm{sec}^{-1}}{\beta_{\chi n}} \left(\frac{P}{1\,\mathrm{GW}}\right)
	\left(\frac{\sigma_{\chi n}}{10^{-28}\,\mathrm{cm}^2}\right)\left(\frac{n_\chi/n_p}{10^{-10}}\right)
	\int_{E_\chi/\beta_{\chi n}}^{\infty}dE_n \frac{1}{N_n}\frac{dN_n}{dE_n}
	\int_{E_\chi/\beta_{\chi n}}^{E_n}\frac{dE}{E^2}\frac{10^{-24}\,\mathrm{cm}^2}{\sigma_{pn}(E)},
\end{align}
where we assume that the proton number density $n_p$ is constant inside the reactor.
\subsection{Event rate at the detector}
In the previous subsection, we calculated the accelerated DM spectrum at the source.
To calculate the event rate at the detector, we need to include DM slow-down in the shielding between the reactor and the detector. In a realistic situation, one would have to take into account reactor walls, concrete shielding around the reactor, as well as shielding surrounding the detector. Here we pursue a simplified approach, simply lumping all amount of shielding into a single parameter characterizing its overall thickness.

We start from the event rate of the DM reaching the detector without being attenuated by the shielding (zero collision case):
\begin{align}
	\left[\frac{d N_\mathrm{eve}}{dE_\mathrm{eve}}\right]_{0}
	= \int dE_\chi \frac{d^2 N_\chi}{dt dE_\chi}
	\times 
	\frac{d\sigma_{\chi t}}{dE_\mathrm{eve}}(E_\chi, E_\mathrm{eve})
	\times \frac{t_\mathrm{det} V_\mathrm{det} n_t}{4\pi l_\mathrm{det}^2}
	\times e^{-n_s \sigma_{\chi s} l_s},
\end{align}
where $V_\mathrm{det}$ is the volume of the detector,
$n_t$ is the number density of the target particles $t$ inside the detector,
$t_\mathrm{det}$ is the exposure time,
$l_\mathrm{det}$ is the distance between the detector and the reactor, 
$d\sigma_{\chi t}/dE_\mathrm{eve}$ is the differential cross section
between the DM and the target particles,
and the subscript ``$0$" stands for zero-scattering.
We assume that the shielding has the thickness $l_s$, the number density $n_s$,
and the scattering cross section $\sigma_{\chi s}$.
For the isotropic scattering between $\chi$ and $t$, we obtain
\begin{align}
	\left[\frac{dN_\mathrm{eve}}{dE_\mathrm{eve}}\right]_{0}
	= \frac{n_t \sigma_{\chi t}V_\mathrm{det} t_\mathrm{det}}{4\pi \beta_{\chi t} l_\mathrm{det}^2}
	e^{-n_s \sigma_{\chi s} l_s}
	\int_{E_\mathrm{eve}/\beta_{\chi t}}^{\infty} \frac{dE_\chi}{E_\chi} \frac{d^2 N_\chi}{dt dE_\chi},
\end{align}
where $\beta_{\chi t} = 4m_\chi m_t/(m_\chi + m_t)^2$.

When the cross section becomes sizable, 
the probability of having zero-scattering is exponentially suppressed
and multiple-scattering contributions become significant.
Since the DM is significantly lighter than the particles inside the shielding in the case of our interest, $m_\chi/m_s \lesssim 0.1$,
the DM particles retain sufficient kinetic energy even after experiencing multiple scatterings
(see App.~\ref{app:energy_loss_fn} for more details).
To calculate the multiple-scattering contributions properly, we perform a Monte Carlo simulation.
For simplicity, we assume that the shielding is spherical around the reactor core (whose finite size we ignore) with the radius $l_s$,
and that the shielding is composed solely of Si with 95\,\% of $A= 28$ and 5\,\% of $A=29$, 
and its density $2.4\,\mathrm{g}/\mathrm{cm}^3$
(the typical value of concrete).
For the SI case, we then have
\begin{align}
	&n_s = 5.2\times 10^{22}\,\mathrm{cm}^{-3},
	\quad
	\sigma_{\chi s}^{\mathrm{SI}} = 
	\mathrm{min}\left[\sigma_{\chi n}^{\mathrm{SI}}\times 28^2 \times
	\left(\frac{m_{\mathrm{Si}}(m_\chi+m_n)}{m_n (m_\chi + m_\mathrm{Si})}\right)^2,
	4\pi r_\mathrm{Si}^2
	\right],
\end{align}
with $m_\mathrm{Si} = 28m_n$ and $r_\mathrm{Si} = 4.03 \times 10^{-13}\,\mathrm{cm}$. For elastic cross sections, we use an ansatz that interpolates between a perturbative answer where nucleons of a given nucleus contribute to the scattering amplitude, and the strong coupling regime when the scattering occurs mostly at the nuclear surface.
In the SD case,  the shielding comes dominantly from ${}^{29}$Si with the nuclear spin $I = 1/2$. 
We then take
\begin{align}
	&n_s = 2.6\times 10^{21}\,\mathrm{cm}^{-3},
	\quad
	\sigma_{\chi s}^{\mathrm{SD}} = 
	\sigma_{\chi n}^{\mathrm{SD}}
	\times \left(\frac{m_{\mathrm{Si}}(m_\chi+m_n)}{m_n (m_\chi + m_\mathrm{Si})}\right)^2 \times \frac{4(I+1)}{3I}
	\times \left(\langle S_n \rangle  + \langle S_p \rangle\right)^2,
	\label{eq:sigmaSD_rescale}
\end{align}
where we assume the isospin symmetric coupling and use the spin average value of the neutron and proton 
$\langle S_n \rangle + \langle S_p \rangle = 0.172$ quoted in~\cite{Klos:2013rwa,Bramante:2022pmn}.

Throughout our computation, the DM is non-relativistic and the target particle is at rest in the laboratory frame.
The scattering length is then given by $l_\mathrm{coll} = 1/n_s \sigma_{\chi s}$.
In our Monte Carlo simulation, we take the step size as 
$\Delta l = \mathrm{min}\left(0.1\times l_\mathrm{coll},~0.01\times l_s\right)$.
For each step, we generate a random variable uniformly distributing between 0 and 1.
If it is smaller than $1 - e^{-\Delta l/l_\mathrm{coll}}$,
we generate a scattering and change the direction of the DM particle isotropically.
We stop each Monte Carlo step once the particle goes outside the sphere of radius $l_s$,
and in each process, we keep track of the number of scatterings that the DM experienced.
Note that the numerical result depends only on the ratio $l_\mathrm{coll}/l_s$.
\begin{figure}[t]
	\centering
	\includegraphics[width=0.475\linewidth]{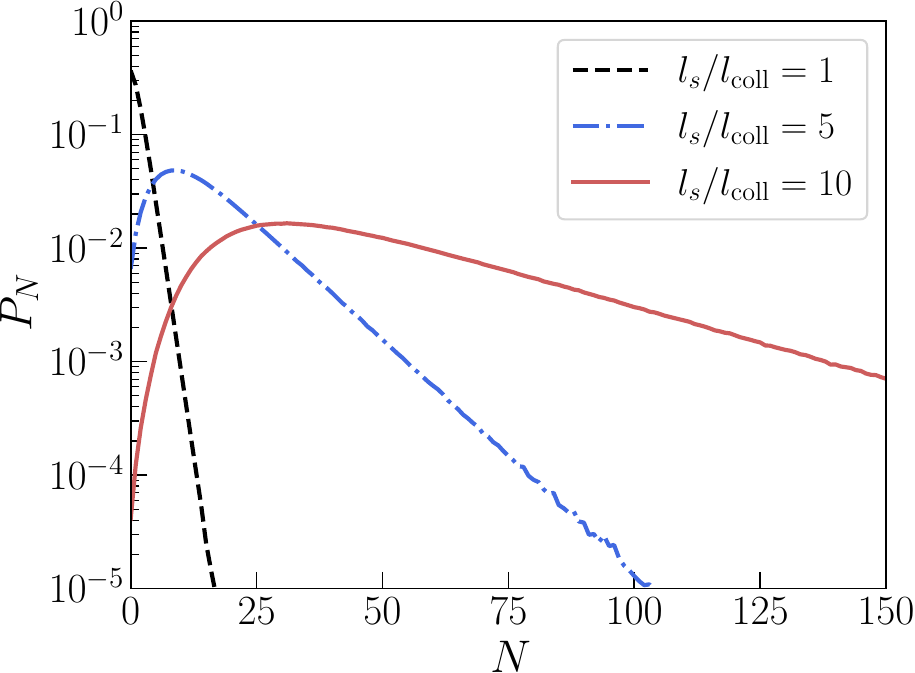}
	\hspace{2.5mm}
	\includegraphics[width=0.475\linewidth]{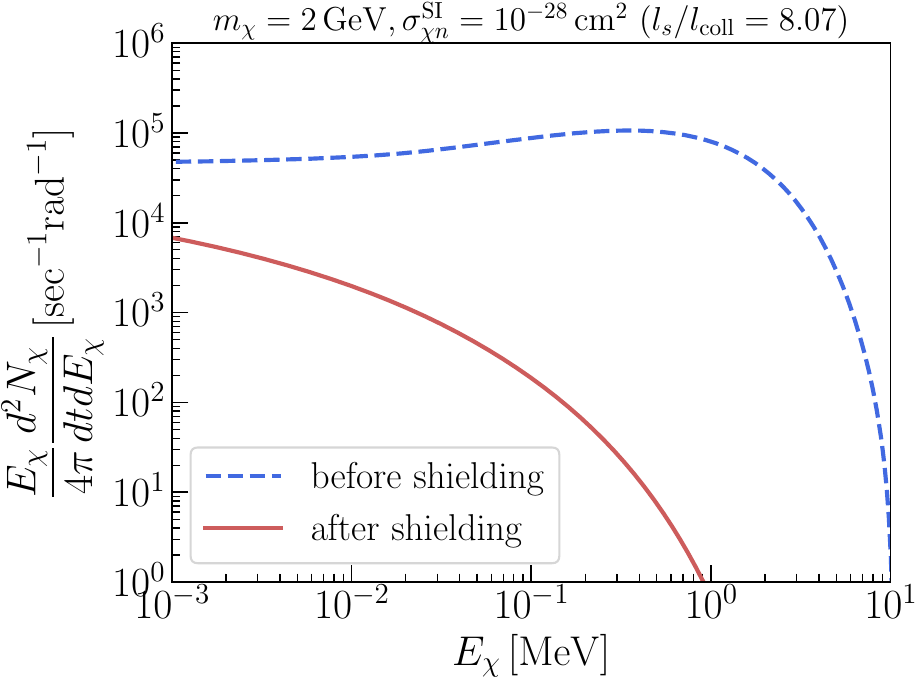}
	\caption{ 
		\emph{Left panel:} The probability of DM particles being scattered $N$ times before exiting the shielding,
		$P_N$, for three different values of $l_s/l_\mathrm{coll}$, computed with our Monte Carlo simulation.
		In this plot, we generated $10^{7}$ events and the wiggles that show up around 
		$P_N \sim 10^{-4} - 10^{-5}$ are due to statistical fluctuations.
		\emph{Right panel:} The DM energy spectrum emitted from the nuclear reactor 
		before and after $5\,\mathrm{m}$ of the shielding in the SI case.
		The parameters are taken as $P = 3.9\,\mathrm{GW}$, $m_\chi = 2\,\mathrm{GeV}$,
		and $\sigma_{\chi n}^{\mathrm{SI}} = 10^{-28}\,\mathrm{cm}^2$,
		corresponding to $l_s/l_\mathrm{coll} = 8.07$.
	}
	\label{fig:Pn_DMflux}
\end{figure}
In the left panel of Fig.~\ref{fig:Pn_DMflux}, 
we show the probability of a DM particle being scattered $N$ times before exiting the shielding, $P_N$, 
computed by our Monte Carlo simulation for different values of $l_s/l_\mathrm{coll}$. 
It is clear from the plot that once $l_s/l_\mathrm{coll}$ exceeds unity, 
the probability of experiencing multiple times of scatterings increases drastically.
Eventually we expect that the average number of scatterings before exiting the shielding 
scales as $(l_s/l_\mathrm{coll})^2$ due to the random walk process.
The DM spectrum after the shielding is given by
\begin{align}
	\left.\frac{dN_\chi}{dE_\chi}\right\vert_\mathrm{shield}
	= \int dE\frac{dN_\chi}{dE_\chi}(E) \times \sum_{N=0}^{\infty} P_N \times f_N(E, E_\chi),
\end{align}
where $f_N$ is the probability of the DM with the initial kinetic energy $E$ ending up with 
the kinetic energy $E_\chi$ after $N$-scatterings (see App.~\ref{app:energy_loss_fn}).
We may assume for simplicity that the flux is isotropic and all the particles are still radially outgoing 
after the shielding.
The full event rate is then given by
\begin{align}
	\frac{d N_\mathrm{eve}}{dE_\mathrm{eve}}
	= \int dE_\chi \left.\frac{d^2 N_\chi}{dt dE_\chi}\right\vert_\mathrm{shield}
	\times 
	\frac{d\sigma_{\chi t}}{dE_\mathrm{eve}}(E_\chi, E_\mathrm{eve})
	\times \frac{t_\mathrm{det} V_\mathrm{det} n_t}{4\pi l_\mathrm{det}^2}.
\end{align}
Here, note that, the subscript ``eve" refers to quantities at the location of the detector.
To see the effects of multiple scatterings, we plot the DM spectrum emitted from the reactor
before and after the shielding, with $l_s = 5\,\mathrm{m}$, in the SI case
in the right panel of Fig.~\ref{fig:Pn_DMflux}.
We take $P = 3.9\,\mathrm{GW}$, 
$m_\chi = 2\,\mathrm{GeV}$, and $\sigma_{\chi n}^{\mathrm{SI}} = 10^{-28}\,\mathrm{cm}^2$, 
which leads to $l_s/l_\mathrm{coll} = 8.07$. 
Although most DM particles experience multiple scatterings and the DM spectrum
is depleted, a small portion of DM particles still retains sufficient kinetic energy, primarily due to the scattering kinematics.
Since the initial energy of DM particles is high enough compared to the threshold energy of 
the CE$\nu$NS experiments,
the multiple scattering components greatly enhance the sensitivity of the neutrino experiments
to our scenario, as we will see in the next section.

\section{Constraints from existing nuclear reactors}
\label{sec:constraints}
Recently CE$\nu$NS has attracted a lot of attention,
and extensive experimental programs world-wide aim at 
detecting the CE$\nu$NS signals both from neutrinos produced at fixed target experiments and 
from the reactor neutrinos. 
The latter type of experiment locates their detectors close to nuclear reactors, 
and therefore provides an ideal platform for probing the Earth-bound DM accelerated by the reactor neutrons.
In the following, we focus on the CONUS experiment~\cite{CONUS:2020skt,CONUS:2021dwh} as a representative
of such an experiment.
Other experiments that are/will be sensitive to the accelerated Earth-bound DM  include Bullkid~\cite{Colantoni:2020cet}, CONNIE~\cite{CONNIE:2019swq}, 
Dresden-I\hspace{-0.25mm}I~\cite{Colaresi:2021kus,Colaresi:2022obx}, 
MINER~\cite{MINER:2016igy}, NEON~\cite{NEON:2022hbk}, $\nu$-cleus~\cite{Strauss:2017cuu}, 
$\nu$GeN~\cite{nGeN:2022uje}, RED-100~\cite{Akimov:2017hee}, Ricochet~\cite{Billard:2016giu}, 
Texono~\cite{Wong:2015kgl}, SBC~\cite{SBC:2021yal},
among others.

The CONUS experiment~\cite{CONUS:2020skt,CONUS:2021dwh} in Brokdorf, Germany
searches for the CE$\nu$NS events from reactor neutrinos.
The commercial power plant in Brokdorf has the thermal power $P = 3.9\,\mathrm{GW}$.
The CONUS detector is composed of Ge (for simplicity we assume 92\,\% of ${}^{74}$Ge and 8\,\% of ${}^{73}$Ge)
and is located $l_\mathrm{det} = 17.1\,\mathrm{m}$ away from the reactor core center,
with shielding of $l_s = 5\,\mathrm{m}$ composed of concrete.\footnote{
	The CONUS experiment has an additional shielding around the detectors,
	which we ignore in our analysis for simplicity.

}
We may use the analysis in~\cite{CONUS:2020skt} with the dataset of 248.7 kg$\times$day,
resulting in $n_t V_\mathrm{det} t_\mathrm{det} = 1.75\times 10^{32}\,\mathrm{sec}$
for the SI case and $1.4\times 10^{31}\,\mathrm{sec}$ for the SD case (originating from ${}^{73}$Ge), respectively.
For $\sigma_{\chi t}$, the scattering cross section on target nuclei, we use
\begin{align}
	\sigma_{\chi t}^{\mathrm{SI}} = 
	\mathrm{min}\left[\sigma_{\chi n}^{\mathrm{SI}}\times 74^2 \times
	\left(\frac{m_{\mathrm{Ge}}(m_\chi+m_n)}{m_n (m_\chi + m_\mathrm{Ge})}\right)^2,
	4\pi r_\mathrm{Ge}^2
	\right],
\end{align}
with $m_\mathrm{Ge} = 74 m_n$ and $r_\mathrm{Ge} = 5.26 \times 10^{-13}\,\mathrm{cm}$ for the SI case,
and  
\begin{align}
	\sigma_{\chi t}^{\mathrm{SD}} = 
	\sigma_{\chi n}^{\mathrm{SD}}
	\times \left(\frac{m_{\mathrm{Ge}}(m_\chi+m_n)}{m_n (m_\chi + m_\mathrm{Ge})}\right)^2 \times \frac{4(I+1)}{3I}
	\times \left(\langle S_n \rangle  + \langle S_p \rangle\right)^2,
\end{align}
with $I = 9/2$ and $\langle S_n \rangle  + \langle S_p \rangle = 0.47$~\cite{Klos:2013rwa,Bramante:2022pmn} 
for the SD case, respectively.
Again, the formula in the SI case interpolates between perturbative and non-perturbative regimes. (A more accurate treatment can be implemented by explicitly introducing the nucleon-DM potential, and calculating the resulting scattering cross section.)

In our case, the DM scatters off the nuclei, not electrons, inside the detector,
and hence we need a conversion of the nuclear recoil energy to the electron-equivalent recoil energy (which is better calibrated).
The quenching factor $k$, defined as the ratio of the ionization energy generated by nuclear recoils
to the one produced by electron recoils with the same energy transfer, is rather uncertain in the energy range of our interest.
We may take $k = 0.18$~\cite{CONUS:2020skt} and assume that it is energy-independent for simplicity.
The energy region of interest depends both on the detectors and different scientific runs in the analysis in~\cite{CONUS:2020skt},
but we may take it as $[0.3\,\mathrm{keV}_{ee}, 1\,\mathrm{keV}_{ee}]$ to simplify our analysis,
where the subscript ``$ee$" stands for the electron-equivalent.
Assuming that the collected charge output is linear in the electron recoil energy, we thus require~\cite{CONUS:2020skt}
\begin{align}
	\int_{1.7\,\mathrm{keV}}^{5.6\,\mathrm{keV}} dE_\mathrm{eve} \frac{d N_\mathrm{eve}}{dE_\mathrm{eve}}
	< 85,
\end{align}
to derive the constraints on the Earth-bound DM from the CONUS experiment.
Note that, even though the CONUS experiment focuses on $\mathcal{O}(1)\,\mathrm{keV}$ recoil energy since
their main target is the CE$\nu$NS events, our DM flux is extended up to $\mathcal{O}(1)\,\mathrm{MeV}$ 
as shown in~\ref{fig:Pn_DMflux}.
Therefore, an analysis dedicated to our DM flux with e.g. higher energy threshold may further improve the sensitivity.

\begin{figure}[t]
	\centering
	\includegraphics[width=0.475\linewidth]{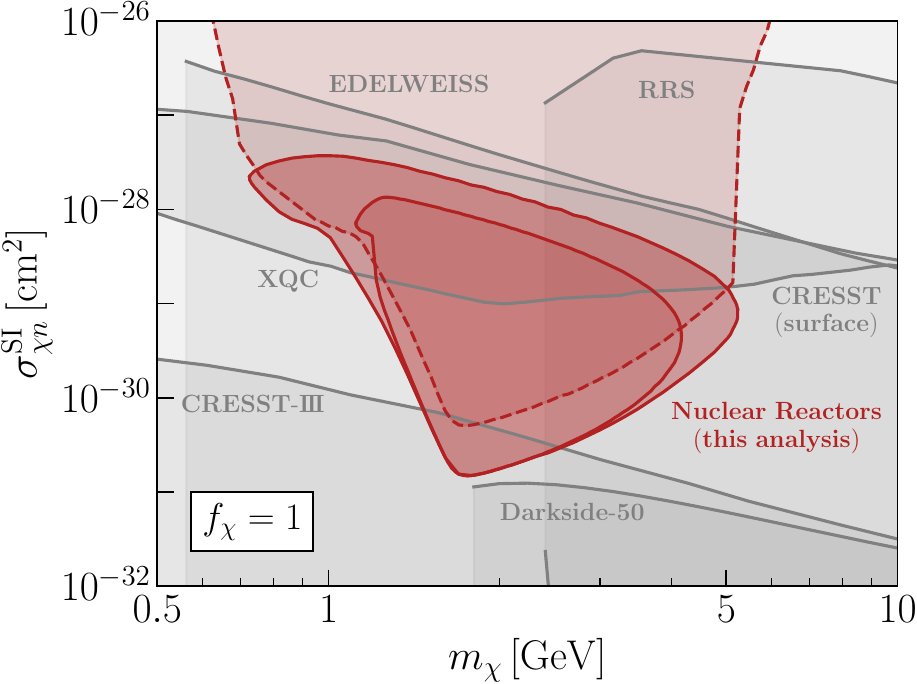}
	\hspace{2.5mm}
	\includegraphics[width=0.475\linewidth]{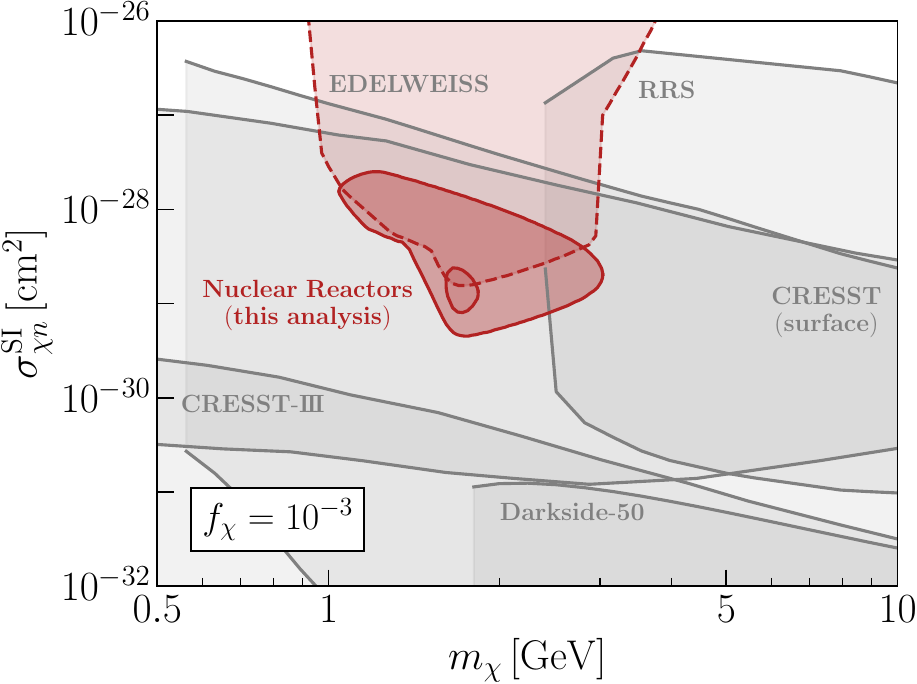}
	\caption{ Constraints on the spin-independent cross section $\sigma_{\chi n}^{\mathrm{SI}}$ 
		from the CONUS experiment~\cite{CONUS:2020skt} 
		for $f_\chi = 1$ (left) and $f_\chi = 10^{-3}$ (right).
		The bigger red shaded regions are the full result with the multiple-scattering components, whereas, the smaller red regions are the result with only the zero scattering component, respectively. Red dashed contours indicate possible reach when DD experiments are placed in proximity of the nuclear research reactors (see text for details). In comparison, the existing surface/underground direct detection constraints from CRESST-I\hspace{-0.25mm}I\hspace{-0.25mm}I~\cite{CRESST:2019jnq}, CRESST (surface)~\cite{CRESST:2017ues}, Darkside-50~\cite{DarkSide:2018bpj}, EDELWEISS (surface)~\cite{EDELWEISS:2019vjv}, rocket based X-ray Quantum Calorimeter (XQC)~\cite{Erickcek:2007jv}, and balloon-based RRS experiment~\cite{Rich:1987st} are shown in gray shaded regions. 	
		Cosmological constraints from CMB, Lyman-$\alpha$ measurements~\cite{Gluscevic:2017ywp,Boddy:2018wzy,Rogers:2021byl} 
		as well as the constraints discussed in~\cite{Neufeld:2018slx}
		do not cover any additional parameter space, and therefore are not shown for clarity.
	}
	\label{fig:constraint_mvssigma}
\end{figure}

Our results in the SI case 
are shown in Fig.~\ref{fig:constraint_mvssigma} for $f_\chi = 1$ (the left panel) and $f_\chi = 10^{-3}$ (the right panel).
The region enclosed by the red line is the full result including the multiple scatterings while 
the smaller region enclosed by the red line is the result with only the zero scattering component.
The figures show that the multiple scattering contributions indeed enhance the sensitivity for the large values of the cross section.
In comparison, we show the direct detection constraints from 
CRESST-I\hspace{-0.25mm}I\hspace{-0.25mm}I~\cite{CRESST:2019jnq}, CRESST (surface)~\cite{CRESST:2017ues},
Darkside-50~\cite{DarkSide:2018bpj}, 
EDELWEISS (surface)~\cite{EDELWEISS:2019vjv}, rocket based X-ray Quantum Calorimeter (XQC)~\cite{Erickcek:2007jv},
and balloon-based RRS experiment~\cite{Rich:1987st}.
We observe that the current parameter region of  the accelerated Earth-bound DM explored by the CONUS experiment
is already covered by the other direct detection experiments,
both for $f_\chi = 1$, and for smaller values of $f_\chi$.

As it is evident from Fig.~\ref{fig:constraint_mvssigma}, the main obstacle for extending the constraints towards larger cross sections
in the SI case is the amount of shielding employed in the searches at commercial reactors. The CONUS experiment loses the sensitivity for $\sigma_{\chi n}^{\mathrm{SI}} \gtrsim 10^{-28}\,\mathrm{cm}^2$ due to the shielding. Research reactors, that typically have two-to-three orders of magnitude less power are not widely used so far for the detection of CE$\nu$NS, where the main limiting factor is the neutrino flux. For the thermalized DM searches, however, the research nuclear reactors may indeed provide additional sensitivity. Assuming that a DD experiment can operate at $\sim 3$\,m distance from the research reactor core (a factor of 10 closer than for a typical placement of a detector at commercial reactor)  with the thermal power of $P = 10\,\mathrm{MW}$, and employ only 1\,m of shielding
\footnote{For instance, 
	the FRJ-1 research reactor at J\"ulich was used to search for (pseudo)scalar particles in~\cite{Faissner:1992ny}
	with minimal amount of shielding, $l_s \sim \mathcal{O}(10)\,\mathrm{cm}$.}, we plot the projected sensitivity with the dashed line contour. One can see that the in $f_\chi=1$ case, the sensitivity extends to the region constrained only by the rocket-based XQC experiment. At $f_\chi =10^{-3}$, new unconstrained parameter space can be covered. 

The new sensitivity that we claim can be achieved is right in the range of masses and cross sections for a hypothetical exotic stable di-nucleon/sexaquark state \cite{Farrar:2022mih}. If stable, such particles may contribute to the DM abundance. The upscattering in the nuclear research reactors is perhaps one of the most promising ways to limit/study such model. 

\begin{figure}[t]
	\centering
	\includegraphics[width=0.475\linewidth]{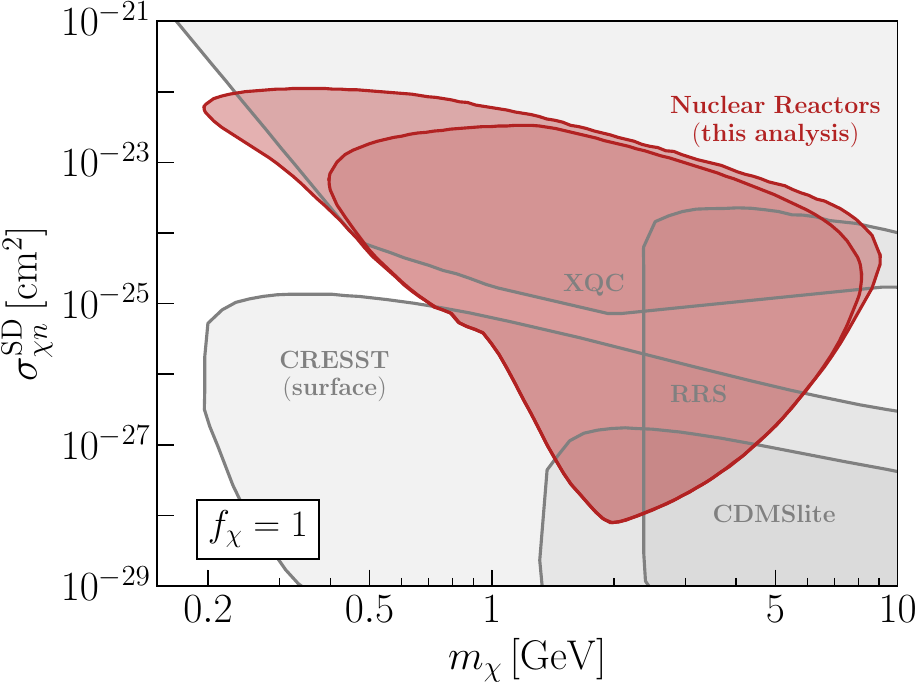}
	\hspace{2.5mm}
	\includegraphics[width=0.475\linewidth]{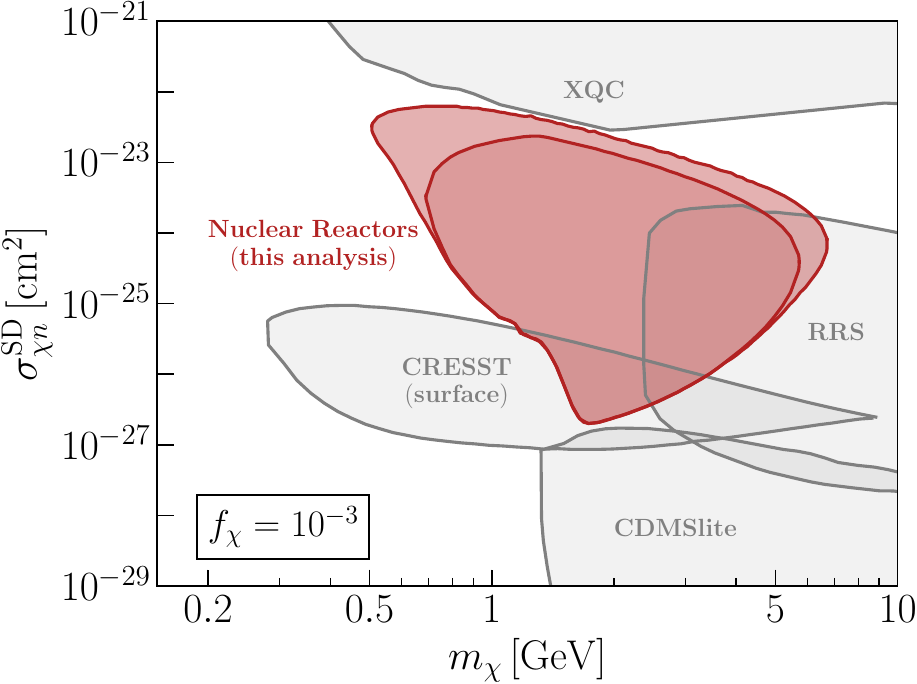}
	\caption{ 
		Constraints on the spin-dependent cross section $\sigma_{\chi n}^{\mathrm{SD}}$ from the CONUS experiment~\cite{CONUS:2020skt} for $f_\chi = 1$ (left) and $f_\chi = 10^{-3}$ (right), including the full multiple scattering contributions (the bigger red-shaded region) and the zero-scattering contribution only (the smaller red-shaded region). In comparison, the existing surface/underground direct detection constraints from CDMSlite~\cite{SuperCDMS:2017nns}, CRESST (surface)~\cite{CRESST:2017ues}, rocket based X-ray Quantum Calorimeter (XQC)~\cite{Erickcek:2007jv}, and balloon-based RRS experiment~\cite{Rich:1987st} (adapted from~\cite{Bramante:2022pmn}) are shown in gray shaded regions. Cosmological constraints from CMB, Lyman-$\alpha$ measurements~\cite{Gluscevic:2017ywp,Boddy:2018wzy,Rogers:2021byl} (not shown) assumes DM-proton scattering, and can be alleviated for neutron only scattering.
	}
	\label{fig:constraint_SD}
\end{figure}

Next, our results in the SD case are shown in Fig.~\ref{fig:constraint_SD} for $f_\chi = 1$ (the left panel) and
$f_\chi = 10^{-3}$ (the right panel). The region enclosed by the red line is the constraint from the CONUS experiment,
while the other existing DD experiments are shown in gray. 
In this case, we see that the CONUS experiment probes a new parameter region, not covered
by the existing DD experiments.
Note that the atmosphere is composed mainly of ${}^{14}$N which has a nuclear spin one,\footnote{
	Refs.~\cite{Hooper:2018bfw,EDELWEISS:2019vjv,Bramante:2022pmn} 
	quoted $\langle S_n \rangle = \langle S_p\rangle = 1/2$ for ${}^{14}$N. 
	We note that a simple nuclear shell model instead predicts
	$\langle S_n \rangle = \langle S_p \rangle = -1/6$ which reproduces the measured magnetic moment well~\cite{Wolters:1990zz}.
	This enhances the upper bound of the DD experiments, performed on the Earth's surface, shown in Fig.~\ref{fig:constraint_SD}
	by a factor of $9$.
	In our computation of the DM surface density, we adopt the shell model value
	for the cross section with ${}^{14}$N.
}
while the dominant composition of the shielding in the CONUS experiment do not possess a finite nuclear spin.
This indicates that the shielding to the DM particles in the SD case
is effectively smaller, relative to the DD experiments at the Earth's surface, compared to the SI case.
We note that our lower bound scales as $f_\chi^{-1/2}$ 
since the event rate scales as $f_\chi \times \sigma_{\chi n}^2$
where one factor of $\sigma_{\chi n}$ comes from the up-scattering by fast neutrons inside the reactor,
while the other factor of $\sigma_{\chi n}$ is from the scattering inside the detector.
On the other hand, the bounds from the DD experiments scale as $f_\chi^{-1}$,
therefore the parameter region newly explored by the CONUS experiment further enlarges 
for smaller values of $f_\chi$ than shown in Fig.~\ref{fig:constraint_SD}.

We note several possible avenues for future improvement.
The big effort for detecting CE$v$NS at nuclear reactors is not optimized for searches of the Earth-bound DM.
Therefore, given the extensive programs of the CE$\nu$NS experiments and with a further analysis of the events,
we expect an improvement in the sensitivity in the near future. Depending on the thickness of shielding relative to scattering length, the recoil spectrum from DM could be considerably harder than the CE$\nu$NS recoil spectrum. This is because the momentum transfer to a nucleus in the detector, in case of CE$\nu$NS, is limited by neutrino energy $q\sim E_\nu$, and for accelerated DM scattering, the same recoil momentum can reach $\mathcal{O}(\sqrt{m_p E_n})$. Since reactors produce neutrinos and neutrons in the same energy range, the nuclear recoil energy in the detector for DM scattering compared to CE$\nu$NS can be enhanced by up to $\mathcal{O}(m_n/E_\nu)$ factor. 

We also would like to point out that currently, there is some tension between the results of CE$\nu$NS searches with Dresden-I\hspace{-0.25mm}I \cite{Colaresi:2021kus,Colaresi:2022obx} and the rest of efforts, and especially CONUS. The results of \cite{Colaresi:2021kus,Colaresi:2022obx} are interpreted by authors as CE$\nu$NS, albeit with the use of a somewhat large value for the quenching factor ({\em i.e.} assuming more efficient translation of nuclear recoil into eVee.) This experiment has some residual neutron background, that is absent in CONUS due to more extensive shielding in the latter case. We point out that while these claimed results are difficult to reconcile within the minimal CE$\nu$NS assumptions, the strongly interacting and accelerated DM scattering hypothesis may be consistent with both. Indeed, higher amount of shielding at CONUS may effectively extinguish the DM scattering, while Dresden-I\hspace{-0.25mm}I may still be sensitive to it. This possibility may require a dedicated investigation by the experimental collaborations, with more realistic simulations of DM  moderation than those performed in this paper. 

\begin{figure}[t]
	\centering
	\includegraphics[width=0.45\linewidth]{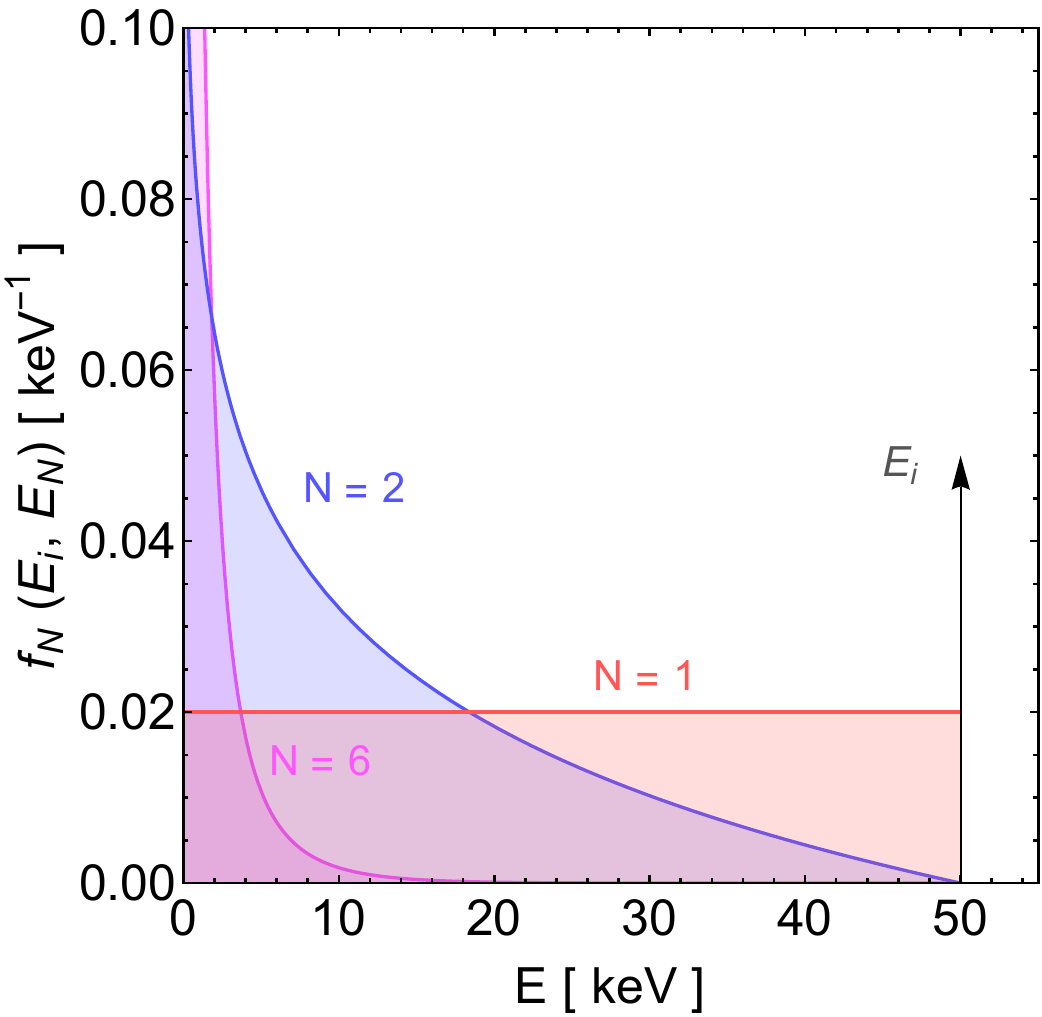}
	\hspace{2.5mm}
	\includegraphics[width=0.45\linewidth]{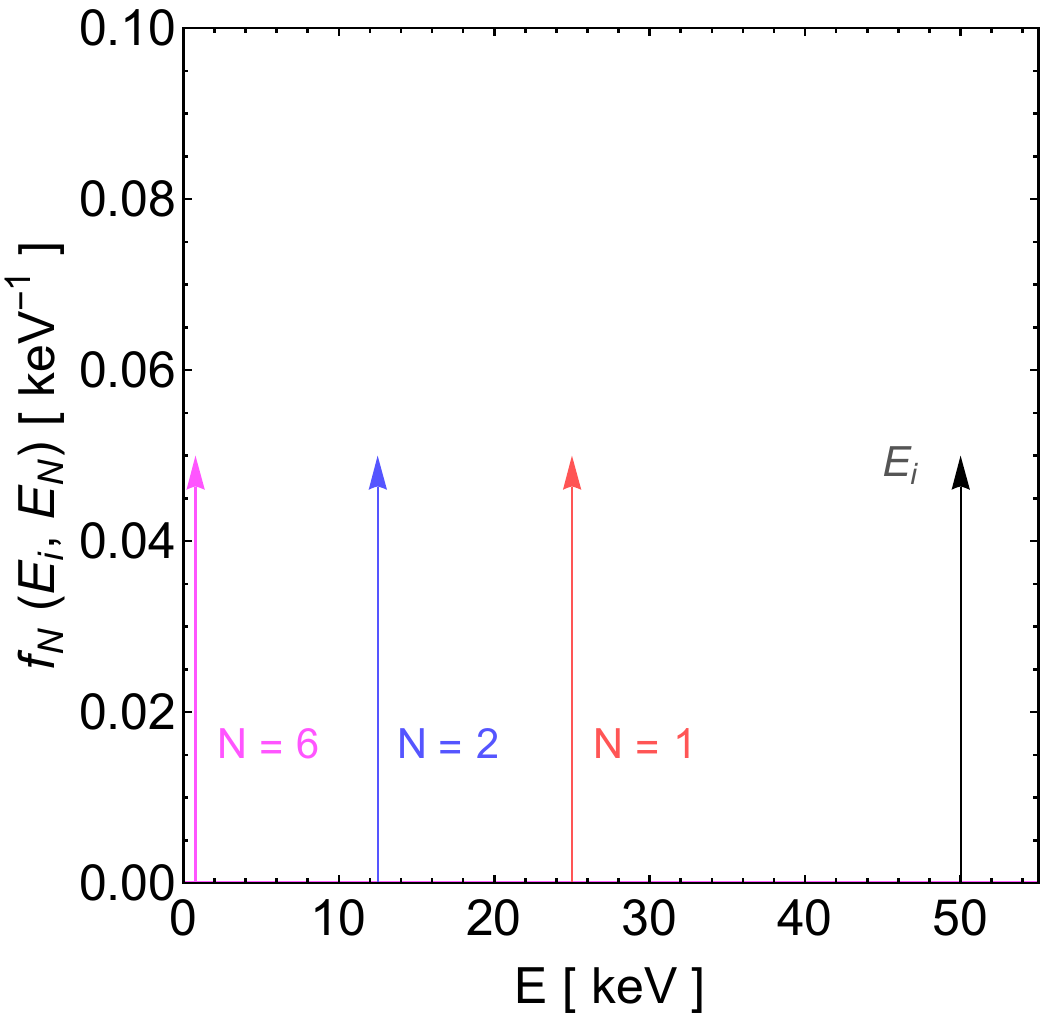}
	\caption{Probability distribution function of final kinetic energy $E_N$ after $N$ number of collisions for the resonant scattering (DM mass = target mass) scenario. (Left) probability distribution of $z$ is used, (Right) $z$ is replaced by its average value $1/2$.}
	\label{fig:distribution}
\end{figure}

\section{Summary}
\label{sec:summary}

We have investigated the use of DD experiments near nuclear reactors in the novel regime. We regard nuclear reactors as $\mathcal{O}({\rm MeV})$ energy neutron beam dumps that have intensities of up to $10^{20}$ neutrons per second. This is a much higher intensity that can be achieved with {\em e.g.} any proton beam of the same energy. Therefore, one can consider new physics scenarios, where nuclear reactors can provide additional sensitivity. 

This paper is dedicated to an investigation of neutron-induced upscattering of dark matter particles with subsequent scattering in a nearby DM detector. Coupling of nuclear reactors and DD experiments is not a hypothetical scheme, but a series of concrete experiments aiming to detect the elastic scattering of antineutrinos. We show that the same experiments provide complementary sensitivity to strongly-interacting dark matter that may accumulate in large quantities inside the Earth in the spin-independent interaction case.
Moreover, in the spin-dependent interaction case, we find that these experiments probe a new parameter region not covered by
the other existing DD experiments.

We found that the thickness of shielding employed in a typical experiment at commercial nuclear reactors searching for CE$\nu$NS is a limiting factor that does not allow for an efficient transmission of neutron-upscattered DM from a reactor to a detector,
especially in the spin-independent interaction case. We point out that the DD experiments at research reactors that require far less shielding may be used to extend sensitivity to so far unconstrained parts of the parameter space.


\paragraph{Acknowledgements}
We would like to thank Dr. M. Lindner for the discussion and explanations of the CONUS experiment setup. 
Y.E. and M.P. are supported in part by U.S. Department of Energy Grant No. desc0011842. A.R. acknowledges support from the National Science Foundation (Grant No.PHY-2020275), and to the Heising-Simons Foundation
(Grant No. 2017-228).

\appendix

\section{Energy Loss Function}
\label{app:energy_loss_fn}
For DM-nucleon scattering, the fractional loss in kinetic energy is given by
\begin{equation}
	\frac{\Delta E}{E} = \frac{4 m_{\chi} m_n}{(m_{\chi}+m_n)^2}z = \beta z\,,
\end{equation}
where $z=\sin^2 (\theta_{\rm CM}/2)$ with $z \in [0,1]$. So, if $E_i$ be the initial kinetic energy of the DM particle and $E_N$ be the final kinetic energy after $N$ collisions, they are related via
\begin{equation}
	E_N = E_i (1-\beta z_1)(1- \beta z_2)...(1-\beta z_N)\,,
\end{equation}
where $z_1,z_2,...,z_N$ are all random variables which have been assumed to be uniformly distributed, \textit{i.e.}, 
their probability distribution function $p(z_i) =1$ (for interactions mediated via heavy mediators, the assumption is perfectly valid~\cite{Dasgupta:2020dik}).
For resonant scattering, \textit{i.e.}, $m_{\chi} = m_n (\beta = 1)$, $E_N$ is an uniformly-distributed random variable which can vary in between 0 to $E_i$, and its probability distribution is simply given by~\cite{Springer1979AlgebraofRandom}
\begin{equation}
	f_N(E_i, E_N) = \frac{1}{E_i} \frac{\log [E_i/E_N]^{N-1}}{(N-1)!}\,.
\end{equation}
For general $\beta$, $E_N$ varies uniformly in between $E_i (1-\beta)^N$ to $E_i$, and its probability distribution is given by~\cite{DETTMANN20092501}
\begin{align}
	f_N(E_i,E_N) = \frac{1}{E_i}\sum_{j=0}^{N-k}& \frac{(-1)^j}{\beta^N (N-1)!} \frac{N!}{(N-j)! j!} \left(\frac{\log (1-\beta)^j}{x}\right)^{N-1}
	\nonumber \\
	&\times \Theta \left((1-\beta)^{N-k}-x\right) \times \Theta \left(x-(1-\beta)^{N+1-k}\right),
\end{align}
where $x= E_N/E_i$ and $k=1,2,..., N$ with $k\leq N$.

In Fig.~\ref{fig:distribution}, we show the probability distribution for the resonant scattering case $(\beta=1)$ for 
$N=1,2,$ and 6 collisions, and we compare it with a simplified approach 
where the random variable $z$ is replaced by 1/2.  
As a concrete example, suppose, we take $``x"$ number of DM particles with a monochromatic initial energy of 50 keV and count the DM particles whose energy falls above, say, 20 keV. In the $z \to 1/2$ approach, it is evident that after first collision, \textit{all} of the $``x"$ number of DM  particles are above threshold, whereas, after second collision, \textit{none} of them are above threshold (right panel). So, either \textit{all} of them (100\%) or \textit{none} of them (0\%).  However, in the realistic scenario, \textit{always} a non-zero fraction of these DM particles are above threshold and this fraction falls off with larger number of collisions (left panel). For example, after first collision, only 60\% of these particles will be above threshold, whereas, after second collision, only 23\% of these particles will be above threshold, and so on.  Therefore, it is evident that $z \to 1/2$ approach is inaccurate as it either over-estimates or under-estimates the realistic fraction for any number of collisions.
In the realistic approach, 
as shown in Fig.~\ref{fig:Pn_DMflux}, a small but non-zero fraction of DM retains large kinetic energy 
even after $\mathcal{O}(10)$ scatterings, and this component extends the sensitivity 
to the larger cross section, as shown in Fig.~\ref{fig:constraint_mvssigma}.

\bibliographystyle{JHEP}
\bibliography{ref}

\providecommand{\href}[2]{#2}\begingroup\raggedright\begin{thebibliography}{10}

\bibitem{Collar:2018ydf}
J.I.~Collar, \emph{{Search for a nonrelativistic component in the spectrum of
  cosmic rays at Earth}},
  \href{https://doi.org/10.1103/PhysRevD.98.023005}{\emph{Phys. Rev. D}
  {\bfseries 98} (2018) 023005}
  [\href{https://arxiv.org/abs/1805.02646}{{\ttfamily 1805.02646}}].

\bibitem{Bramante:2022pmn}
J.~Bramante, J.~Kumar, G.~Mohlabeng, N.~Raj and N.~Song, \emph{{Light dark
  matter accumulating in planets: Nuclear scattering}},
  \href{https://doi.org/10.1103/PhysRevD.108.063022}{\emph{Phys. Rev. D}
  {\bfseries 108} (2023) 063022}
  [\href{https://arxiv.org/abs/2210.01812}{{\ttfamily 2210.01812}}].

\bibitem{Xu:2020qjk}
X.~Xu and G.R.~Farrar, \emph{{Resonant scattering between dark matter and
  baryons: Revised direct detection and CMB limits}},
  \href{https://doi.org/10.1103/PhysRevD.107.095028}{\emph{Phys. Rev. D}
  {\bfseries 107} (2023) 095028}
  [\href{https://arxiv.org/abs/2101.00142}{{\ttfamily 2101.00142}}].

\bibitem{Neufeld:2018slx}
D.A.~Neufeld, G.R.~Farrar and C.F.~McKee, \emph{{Dark Matter that Interacts
  with Baryons: Density Distribution within the Earth and New Constraints on
  the Interaction Cross-section}},
  \href{https://doi.org/10.3847/1538-4357/aad6a4}{\emph{Astrophys. J.}
  {\bfseries 866} (2018) 111}
  [\href{https://arxiv.org/abs/1805.08794}{{\ttfamily 1805.08794}}].

\bibitem{Xu:2021lmg}
X.~Xu and G.R.~Farrar, \emph{{Constraints on GeV Dark Matter interaction with
  baryons, from a novel Dewar experiment}},
  \href{https://arxiv.org/abs/2112.00707}{{\ttfamily 2112.00707}}.

\bibitem{Farrar:2022mih}
G.R.~Farrar, \emph{{A Stable Sexaquark: Overview and Discovery Strategies}},
  \href{https://arxiv.org/abs/2201.01334}{{\ttfamily 2201.01334}}.

\bibitem{Bringmann:2018cvk}
T.~Bringmann and M.~Pospelov, \emph{{Novel direct detection constraints on
  light dark matter}},
  \href{https://doi.org/10.1103/PhysRevLett.122.171801}{\emph{Phys. Rev. Lett.}
  {\bfseries 122} (2019) 171801}
  [\href{https://arxiv.org/abs/1810.10543}{{\ttfamily 1810.10543}}].

\bibitem{Ema:2018bih}
Y.~Ema, F.~Sala and R.~Sato, \emph{{Light Dark Matter at Neutrino
  Experiments}},
  \href{https://doi.org/10.1103/PhysRevLett.122.181802}{\emph{Phys. Rev. Lett.}
  {\bfseries 122} (2019) 181802}
  [\href{https://arxiv.org/abs/1811.00520}{{\ttfamily 1811.00520}}].

\bibitem{Cappiello:2019qsw}
C.V.~Cappiello and J.F.~Beacom, \emph{{Strong New Limits on Light Dark Matter
  from Neutrino Experiments}},
  \href{https://doi.org/10.1103/PhysRevD.104.069901}{\emph{Phys. Rev. D}
  {\bfseries 100} (2019) 103011}
  [\href{https://arxiv.org/abs/1906.11283}{{\ttfamily 1906.11283}}].

\bibitem{McKeen:2023ztq}
D.~McKeen, D.E.~Morrissey, M.~Pospelov, H.~Ramani and A.~Ray, \emph{{Dark
  Matter Annihilation inside Large-Volume Neutrino Detectors}},
  \href{https://doi.org/10.1103/PhysRevLett.131.011005}{\emph{Phys. Rev. Lett.}
  {\bfseries 131} (2023) 011005}
  [\href{https://arxiv.org/abs/2303.03416}{{\ttfamily 2303.03416}}].

\bibitem{Pospelov:2023mlz}
M.~Pospelov and A.~Ray, \emph{{Neutrinos from Earth-Bound Dark Matter
  Annihilation}},  \href{https://arxiv.org/abs/2309.10032}{{\ttfamily
  2309.10032}}.

\bibitem{Pospelov:2019vuf}
M.~Pospelov, S.~Rajendran and H.~Ramani, \emph{{Metastable Nuclear Isomers as
  Dark Matter Accelerators}},
  \href{https://doi.org/10.1103/PhysRevD.101.055001}{\emph{Phys. Rev. D}
  {\bfseries 101} (2020) 055001}
  [\href{https://arxiv.org/abs/1907.00011}{{\ttfamily 1907.00011}}].

\bibitem{Lehnert:2019tuw}
B.~Lehnert, H.~Ramani, M.~Hult, G.~Lutter, M.~Pospelov, S.~Rajendran et~al.,
  \emph{{Search for Dark Matter Induced Deexcitation of $^{180}$Ta$\rm ^m$}},
  \href{https://doi.org/10.1103/PhysRevLett.124.181802}{\emph{Phys. Rev. Lett.}
  {\bfseries 124} (2020) 181802}
  [\href{https://arxiv.org/abs/1911.07865}{{\ttfamily 1911.07865}}].

\bibitem{Majorana:2023ecz}
{\scshape Majorana} collaboration, \emph{{Constraints on the Decay of Ta180m}},
  \href{https://doi.org/10.1103/PhysRevLett.131.152501}{\emph{Phys. Rev. Lett.}
  {\bfseries 131} (2023) 152501}
  [\href{https://arxiv.org/abs/2306.01965}{{\ttfamily 2306.01965}}].

\bibitem{Das:2022srn}
A.~Das, N.~Kurinsky and R.K.~Leane, \emph{{Dark Matter Induced Power in Quantum
  Devices}}, \href{https://doi.org/10.1103/PhysRevLett.132.121801}{\emph{Phys.
  Rev. Lett.} {\bfseries 132} (2024) 121801}
  [\href{https://arxiv.org/abs/2210.09313}{{\ttfamily 2210.09313}}].

\bibitem{Pospelov:2020ktu}
M.~Pospelov and H.~Ramani, \emph{{Earth-bound millicharge relics}},
  \href{https://doi.org/10.1103/PhysRevD.103.115031}{\emph{Phys. Rev. D}
  {\bfseries 103} (2021) 115031}
  [\href{https://arxiv.org/abs/2012.03957}{{\ttfamily 2012.03957}}].

\bibitem{McKeen:2022poo}
D.~McKeen, M.~Moore, D.E.~Morrissey, M.~Pospelov and H.~Ramani,
  \emph{{Accelerating Earth-bound dark matter}},
  \href{https://doi.org/10.1103/PhysRevD.106.035011}{\emph{Phys. Rev. D}
  {\bfseries 106} (2022) 035011}
  [\href{https://arxiv.org/abs/2202.08840}{{\ttfamily 2202.08840}}].

\bibitem{CHANDLER:2022gvg}
{\scshape CHANDLER, CONNIE, CONUS, Daya Bay, JUNO, MTAS, NEOS, NuLat, PROSPECT,
  RENO, Ricochet, ROADSTR Near-Field Working Group, SoLid, Stereo,
  Valencia-Nantes TAGS, vIOLETA, WATCHMAN} collaboration, \emph{{High Energy
  Physics Opportunities Using Reactor Antineutrinos}},
  \href{https://arxiv.org/abs/2203.07214}{{\ttfamily 2203.07214}}.

\bibitem{Abdullah:2022zue}
M.~Abdullah et~al., \emph{{Coherent elastic neutrino-nucleus scattering:
  Terrestrial and astrophysical applications}},
  \href{https://arxiv.org/abs/2203.07361}{{\ttfamily 2203.07361}}.

\bibitem{Dziewonski:1981xy}
A.M.~Dziewonski and D.L.~Anderson, \emph{{Preliminary reference earth model}},
  \href{https://doi.org/10.1016/0031-9201(81)90046-7}{\emph{Phys. Earth Planet.
  Interiors} {\bfseries 25} (1981) 297}.

\bibitem{Bramante:2019fhi}
J.~Bramante, A.~Buchanan, A.~Goodman and E.~Lodhi, \emph{{Terrestrial and
  Martian Heat Flow Limits on Dark Matter}},
  \href{https://doi.org/10.1103/PhysRevD.101.043001}{\emph{Phys. Rev. D}
  {\bfseries 101} (2020) 043001}
  [\href{https://arxiv.org/abs/1909.11683}{{\ttfamily 1909.11683}}].

\bibitem{2002JGRA..107.1468P}
J.M.~{Picone}, A.E.~{Hedin}, D.P.~{Drob} and A.C.~{Aikin}, \emph{{NRLMSISE-00
  empirical model of the atmosphere: Statistical comparisons and scientific
  issues}}, \href{https://doi.org/10.1029/2002JA009430}{\emph{Journal of
  Geophysical Research (Space Physics)} {\bfseries 107} (2002) 1468}.

\bibitem{Gould:1989hm}
A.~Gould and G.~Raffelt, \emph{{Thermal Conduction by Massive Particles}},
  \href{https://doi.org/10.1086/168568}{\emph{Astrophys. J.} {\bfseries 352}
  (1990) 654}.

\bibitem{Leane:2022hkk}
R.K.~Leane and J.~Smirnov, \emph{{Floating dark matter in celestial bodies}},
  \href{https://doi.org/10.1088/1475-7516/2023/10/057}{\emph{JCAP} {\bfseries
  10} (2023) 057} [\href{https://arxiv.org/abs/2209.09834}{{\ttfamily
  2209.09834}}].

\bibitem{https://doi.org/10.1002/2017JB014723}
Y.~Zhang, T.~Sekine, J.-F.~Lin, H.~He, F.~Liu, M.~Zhang et~al., \emph{Shock
  compression and melting of an fe-ni-si alloy: Implications for the
  temperature profile of the earth's core and the heat flux across the
  core-mantle boundary},
  \href{https://doi.org/https://doi.org/10.1002/2017JB014723}{\emph{Journal of
  Geophysical Research: Solid Earth} {\bfseries 123} (2018) 1314}.

\bibitem{CONUS:2020skt}
{\scshape CONUS} collaboration, \emph{{Constraints on elastic neutrino nucleus
  scattering in the fully coherent regime from the CONUS experiment}},
  \href{https://doi.org/10.1103/PhysRevLett.126.041804}{\emph{Phys. Rev. Lett.}
  {\bfseries 126} (2021) 041804}
  [\href{https://arxiv.org/abs/2011.00210}{{\ttfamily 2011.00210}}].

\bibitem{CONUS:2021dwh}
{\scshape CONUS} collaboration, \emph{{Novel constraints on neutrino physics
  beyond the standard model from the CONUS experiment}},
  \href{https://doi.org/10.1007/JHEP05(2022)085}{\emph{JHEP} {\bfseries 05}
  (2022) 085} [\href{https://arxiv.org/abs/2110.02174}{{\ttfamily
  2110.02174}}].

\bibitem{Cranberg:1956zz}
L.~Cranberg, G.~Frye, N.~Nereson and L.~Rosen, \emph{{Fission Neutron Spectrum
  of U-235}}, \href{https://doi.org/10.1103/PhysRev.103.662}{\emph{Phys. Rev.}
  {\bfseries 103} (1956) 662}.

\bibitem{Klos:2013rwa}
P.~Klos, J.~Men\'endez, D.~Gazit and A.~Schwenk, \emph{{Large-scale nuclear
  structure calculations for spin-dependent WIMP scattering with chiral
  effective field theory currents}},
  \href{https://doi.org/10.1103/PhysRevD.88.083516}{\emph{Phys. Rev. D}
  {\bfseries 88} (2013) 083516}
  [\href{https://arxiv.org/abs/1304.7684}{{\ttfamily 1304.7684}}].

\bibitem{Colantoni:2020cet}
I.~Colantoni et~al., \emph{{BULLKID: BULky and Low-Threshold Kinetic Inductance
  Detectors}}, \href{https://doi.org/10.1007/s10909-020-02408-3}{\emph{J. Low
  Temp. Phys.} {\bfseries 199} (2020) 593}.

\bibitem{CONNIE:2019swq}
{\scshape CONNIE} collaboration, \emph{{Exploring low-energy neutrino physics
  with the Coherent Neutrino Nucleus Interaction Experiment}},
  \href{https://doi.org/10.1103/PhysRevD.100.092005}{\emph{Phys. Rev. D}
  {\bfseries 100} (2019) 092005}
  [\href{https://arxiv.org/abs/1906.02200}{{\ttfamily 1906.02200}}].

\bibitem{Colaresi:2021kus}
J.~Colaresi, J.I.~Collar, T.W.~Hossbach, A.R.L.~Kavner, C.M.~Lewis,
  A.E.~Robinson et~al., \emph{{First results from a search for coherent elastic
  neutrino-nucleus scattering at a reactor site}},
  \href{https://doi.org/10.1103/PhysRevD.104.072003}{\emph{Phys. Rev. D}
  {\bfseries 104} (2021) 072003}
  [\href{https://arxiv.org/abs/2108.02880}{{\ttfamily 2108.02880}}].

\bibitem{Colaresi:2022obx}
J.~Colaresi, J.I.~Collar, T.W.~Hossbach, C.M.~Lewis and K.M.~Yocum,
  \emph{{Measurement of Coherent Elastic Neutrino-Nucleus Scattering from
  Reactor Antineutrinos}},
  \href{https://doi.org/10.1103/PhysRevLett.129.211802}{\emph{Phys. Rev. Lett.}
  {\bfseries 129} (2022) 211802}
  [\href{https://arxiv.org/abs/2202.09672}{{\ttfamily 2202.09672}}].

\bibitem{MINER:2016igy}
{\scshape MINER} collaboration, \emph{{Background Studies for the MINER
  Coherent Neutrino Scattering Reactor Experiment}},
  \href{https://doi.org/10.1016/j.nima.2017.02.024}{\emph{Nucl. Instrum. Meth.
  A} {\bfseries 853} (2017) 53}
  [\href{https://arxiv.org/abs/1609.02066}{{\ttfamily 1609.02066}}].

\bibitem{NEON:2022hbk}
{\scshape NEON} collaboration, \emph{{Exploring coherent elastic
  neutrino-nucleus scattering using reactor electron antineutrinos in the NEON
  experiment}},
  \href{https://doi.org/10.1140/epjc/s10052-023-11352-x}{\emph{Eur. Phys. J. C}
  {\bfseries 83} (2023) 226}
  [\href{https://arxiv.org/abs/2204.06318}{{\ttfamily 2204.06318}}].

\bibitem{Strauss:2017cuu}
R.~Strauss et~al., \emph{{The $\nu$-cleus experiment: A gram-scale
  fiducial-volume cryogenic detector for the first detection of coherent
  neutrino-nucleus scattering}},
  \href{https://doi.org/10.1140/epjc/s10052-017-5068-2}{\emph{Eur. Phys. J. C}
  {\bfseries 77} (2017) 506}
  [\href{https://arxiv.org/abs/1704.04320}{{\ttfamily 1704.04320}}].

\bibitem{nGeN:2022uje}
{\scshape \ensuremath{\nu}GeN} collaboration, \emph{{First results of the
  \ensuremath{\nu}GeN experiment on coherent elastic neutrino-nucleus
  scattering}}, \href{https://doi.org/10.1103/PhysRevD.106.L051101}{\emph{Phys.
  Rev. D} {\bfseries 106} (2022) L051101}
  [\href{https://arxiv.org/abs/2205.04305}{{\ttfamily 2205.04305}}].

\bibitem{Akimov:2017hee}
D.Y.~Akimov et~al., \emph{{Status of the RED-100 experiment}},
  \href{https://doi.org/10.1088/1748-0221/12/06/C06018}{\emph{JINST} {\bfseries
  12} (2017) C06018}.

\bibitem{Billard:2016giu}
J.~Billard et~al., \emph{{Coherent Neutrino Scattering with Low Temperature
  Bolometers at Chooz Reactor Complex}},
  \href{https://doi.org/10.1088/1361-6471/aa83d0}{\emph{J. Phys. G} {\bfseries
  44} (2017) 105101} [\href{https://arxiv.org/abs/1612.09035}{{\ttfamily
  1612.09035}}].

\bibitem{Wong:2015kgl}
H.T.-K.~Wong, \emph{{Taiwan EXperiment On NeutrinO \textemdash{} History and
  Prospects}}, \href{https://doi.org/10.1142/S0217751X18300144}{\emph{The
  Universe} {\bfseries 3} (2015) 22}
  [\href{https://arxiv.org/abs/1608.00306}{{\ttfamily 1608.00306}}].

\bibitem{SBC:2021yal}
{\scshape SBC, CE\ensuremath{\nu}NS Theory Group at IF-UNAM} collaboration,
  \emph{{Physics reach of a low threshold scintillating argon bubble chamber in
  coherent elastic neutrino-nucleus scattering reactor experiments}},
  \href{https://doi.org/10.1103/PhysRevD.103.L091301}{\emph{Phys. Rev. D}
  {\bfseries 103} (2021) L091301}
  [\href{https://arxiv.org/abs/2101.08785}{{\ttfamily 2101.08785}}].

\bibitem{CRESST:2019jnq}
{\scshape CRESST} collaboration, \emph{{First results from the CRESST-III
  low-mass dark matter program}},
  \href{https://doi.org/10.1103/PhysRevD.100.102002}{\emph{Phys. Rev. D}
  {\bfseries 100} (2019) 102002}
  [\href{https://arxiv.org/abs/1904.00498}{{\ttfamily 1904.00498}}].

\bibitem{CRESST:2017ues}
{\scshape CRESST} collaboration, \emph{{Results on MeV-scale dark matter from a
  gram-scale cryogenic calorimeter operated above ground}},
  \href{https://doi.org/10.1140/epjc/s10052-017-5223-9}{\emph{Eur. Phys. J. C}
  {\bfseries 77} (2017) 637}
  [\href{https://arxiv.org/abs/1707.06749}{{\ttfamily 1707.06749}}].

\bibitem{DarkSide:2018bpj}
{\scshape DarkSide} collaboration, \emph{{Low-Mass Dark Matter Search with the
  DarkSide-50 Experiment}},
  \href{https://doi.org/10.1103/PhysRevLett.121.081307}{\emph{Phys. Rev. Lett.}
  {\bfseries 121} (2018) 081307}
  [\href{https://arxiv.org/abs/1802.06994}{{\ttfamily 1802.06994}}].

\bibitem{EDELWEISS:2019vjv}
{\scshape EDELWEISS} collaboration, \emph{{Searching for low-mass dark matter
  particles with a massive Ge bolometer operated above-ground}},
  \href{https://doi.org/10.1103/PhysRevD.99.082003}{\emph{Phys. Rev. D}
  {\bfseries 99} (2019) 082003}
  [\href{https://arxiv.org/abs/1901.03588}{{\ttfamily 1901.03588}}].

\bibitem{Erickcek:2007jv}
A.L.~Erickcek, P.J.~Steinhardt, D.~McCammon and P.C.~McGuire,
  \emph{{Constraints on the Interactions between Dark Matter and Baryons from
  the X-ray Quantum Calorimetry Experiment}},
  \href{https://doi.org/10.1103/PhysRevD.76.042007}{\emph{Phys. Rev. D}
  {\bfseries 76} (2007) 042007}
  [\href{https://arxiv.org/abs/0704.0794}{{\ttfamily 0704.0794}}].

\bibitem{Rich:1987st}
J.~Rich, R.~Rocchia and M.~Spiro, \emph{{A Search for Strongly Interacting Dark
  Matter}}, \href{https://doi.org/10.1016/0370-2693(87)90788-X}{\emph{Phys.
  Lett. B} {\bfseries 194} (1987) 173}.

\bibitem{Gluscevic:2017ywp}
V.~Gluscevic and K.K.~Boddy, \emph{{Constraints on Scattering of
  keV\textendash{}TeV Dark Matter with Protons in the Early Universe}},
  \href{https://doi.org/10.1103/PhysRevLett.121.081301}{\emph{Phys. Rev. Lett.}
  {\bfseries 121} (2018) 081301}
  [\href{https://arxiv.org/abs/1712.07133}{{\ttfamily 1712.07133}}].

\bibitem{Boddy:2018wzy}
K.K.~Boddy, V.~Gluscevic, V.~Poulin, E.D.~Kovetz, M.~Kamionkowski and
  R.~Barkana, \emph{{Critical assessment of CMB limits on dark matter-baryon
  scattering: New treatment of the relative bulk velocity}},
  \href{https://doi.org/10.1103/PhysRevD.98.123506}{\emph{Phys. Rev. D}
  {\bfseries 98} (2018) 123506}
  [\href{https://arxiv.org/abs/1808.00001}{{\ttfamily 1808.00001}}].

\bibitem{Rogers:2021byl}
K.K.~Rogers, C.~Dvorkin and H.V.~Peiris, \emph{{Limits on the Light Dark
  Matter\textendash{}Proton Cross Section from Cosmic Large-Scale Structure}},
  \href{https://doi.org/10.1103/PhysRevLett.128.171301}{\emph{Phys. Rev. Lett.}
  {\bfseries 128} (2022) 171301}
  [\href{https://arxiv.org/abs/2111.10386}{{\ttfamily 2111.10386}}].

\bibitem{Faissner:1992ny}
H.~Faissner, P.~Goettlicher, H.~Matela and D.~Samm, \emph{{Search for scalar
  and pseudoscalar bosons emitted in nuclear decays via their interactions}},
  \href{https://doi.org/10.1007/BF01283546}{\emph{Z. Phys. A} {\bfseries 341}
  (1992) 359}.

\bibitem{SuperCDMS:2017nns}
{\scshape SuperCDMS} collaboration, \emph{{Low-mass dark matter search with
  CDMSlite}}, \href{https://doi.org/10.1103/PhysRevD.97.022002}{\emph{Phys.
  Rev. D} {\bfseries 97} (2018) 022002}
  [\href{https://arxiv.org/abs/1707.01632}{{\ttfamily 1707.01632}}].

\bibitem{Hooper:2018bfw}
D.~Hooper and S.D.~McDermott, \emph{{Robust Constraints and Novel Gamma-Ray
  Signatures of Dark Matter That Interacts Strongly With Nucleons}},
  \href{https://doi.org/10.1103/PhysRevD.97.115006}{\emph{Phys. Rev. D}
  {\bfseries 97} (2018) 115006}
  [\href{https://arxiv.org/abs/1802.03025}{{\ttfamily 1802.03025}}].

\bibitem{Wolters:1990zz}
A.A.~Wolters, A.G.M.~van Hees and P.W.M.~Glaudemans, \emph{{p-shell nuclei in a
  (0+2) homega model space. 2. Results}},
  \href{https://doi.org/10.1103/PhysRevC.42.2062}{\emph{Phys. Rev. C}
  {\bfseries 42} (1990) 2062}.

\bibitem{Dasgupta:2020dik}
B.~Dasgupta, A.~Gupta and A.~Ray, \emph{{Dark matter capture in celestial
  objects: light mediators, self-interactions, and complementarity with direct
  detection}}, \href{https://doi.org/10.1088/1475-7516/2020/10/023}{\emph{JCAP}
  {\bfseries 10} (2020) 023}
  [\href{https://arxiv.org/abs/2006.10773}{{\ttfamily 2006.10773}}].

\bibitem{Springer1979AlgebraofRandom}
M.D.~Springer, \emph{The Algebra of Random Variables}, Wiley series in
  probability and mathematical statistics, John Wiley \& Sons (1979).

\bibitem{DETTMANN20092501}
C.P.~Dettmann and O.~Georgiou, \emph{Product of n independent uniform random
  variables},
  \href{https://doi.org/https://doi.org/10.1016/j.spl.2009.09.004}{\emph{Statistics
  \& Probability Letters} {\bfseries 79} (2009) 2501}.

\end{thebibliography}\endgroup

\end{document}